\documentclass{ws-procs9x6}

\def\st{\scriptstyle}

\def\be{\begin{equation}}
\def\ee{\end{equation}}
\def\ba{\begin{eqnarray}}
\def\ea{\end{eqnarray}}
\def\bsu{\begin{subequations}}
\def\esu{\end{subequations}}

\def\a{\alpha}
\def\b{\beta}
     
\def\d{\delta}

\def\l{\lambda}
\def\m{\mu}
\def\n{\nu}
\def\o{\omega}   \def\O{\Omega}

\def\t{\tau}

\def\la{\label}
\def\pd{\partial}
\def\le{\left}
\def\ri{\right}

\def\mm{{\mathtt f}}

\def\etal {{\it et al.}}

\begin{document}

\title{Equivalence principle in cosmology}

\author{Sergei\ Kopeikin$^*$}

\address{Department of Physics and Astronomy, \\University of Missouri,
Columbia, MO 65211, USA\\
$^*$E-mail: kopeikins@missouri.edu}

\begin{abstract}
We analyse the Einstein equivalence principle (EEP) for a Hubble observer in Friedmann-Lema\^itre-Robertson-Walker (FLRW) spacetime. We show that the affine structure of light cone in the FLRW spacetime should be treated locally in terms of the optical metric $g_{\a\b}$ which is not reduced to the Minkowski metric $\mm_{\a\b}$ due to the non-uniform parametrization of the local equations of light propagation with the proper time of the observer's clock. The physical consequence of this difference is that the Doppler shift of radio waves measured locally, is affected by the Hubble expansion. 
\end{abstract}
\bodymatter
\section{Introduction}

Experimental gravity in the solar system is done under assumption that spacetime is asymptotically-flat\cite{2011rcms.book.....K}. In contrast, cosmology makes use of FLRW metric\cite{weinberg_2008}
\be\la{1}
ds^2=-dt^2+a^2\d_{ij}dy^idy^j\;,
\ee
where $t$ is the cosmological time, $y^i$ are global coordinates, Latin indices $i,j,k,..$ take values $1,2,3$, the scale factor $a=a(t)$ is found from Einstein's equations. We use a geometric system of units in which $G=c=1$. 
Metric (\ref{1}) can be reduced to a conformally-flat metric by introducing a conformal time $\eta=\eta(t)$ defined by 
$dt/a(t)=d\eta$. It brings (\ref{1}) to
\be\la{q1}
ds^2=a^2\mm_{\a\b}dy^\a dy^\b\;,
\ee
where $y^\a=(y^0,y^i)=\le(\eta,y^i\ri)$, $\mm_{\a\b}={\rm diag}(-1,1,1,1)$ is the Minkowski metric. The cosmological time $t$ is physical and can be measured by the Hubble observers with the help of clocks while the conformal time $\eta$ is merely a convenient coordinate parameter\cite{weinberg_2008}. Typically, the cosmological metric (\ref{q1}) is applied to describe the properties of spacetime on cosmological distances. We shall assume that metric (\ref{q1}) describes the background geometry of spacetime on all scales.

\section{Einstein's Equivalence Principle}
The Einstein equivalence principle (EEP) states that at each point on a manifold with an arbitrary gravitational field, it is possible to chose local inertial coordinates such that, within a sufficiently small region of the point in question, the laws of nature take the same form as in non-accelerated Cartesian coordinates\cite{will_1993}. According to EEP each Hubble observer carries out local inertial coordinates (LIC), $x^\a=(x^0,x^i)=(\t,x^i)$ such that the physical metric ${\st\cal G}_{\a\b}=a^2(\eta)\mm_{\a\b}$, given by (\ref{q1}), is reduced to the Minkowski metric, $\mm_{\a\b}$, and the affine connection is nil on the worldline of the observer. EEP also asserts that free fall of electrically-neutral test particles and photons is geodesic and their equations of motion in LIC are given locally by the same equation
\be\la{bq1}
\frac{d^2x^\a}{d{\t}^2}=0\;,
\ee
where $\t$ is the coordinate time of the LIC, with all tidal terms being neglected. EEP is not hold in conformal spacetime as shown in Eq. (6) below. 
\section{Local Inertial Coordinates}
Let us construct LIC in the vicinity of the worldline of a Hubble observer who is located at the origin of LIC, $x^i=0$. The observer carries out an ideal clock that measures the parameter on the observer's worldline which is identified with the cosmic time $t$ in (\ref{1}). EEP suggests that the physical spacetime interval (\ref{q1}) written down in LIC, reads
\be\la{act6}
ds^2=\mm_{\a\b}dx^\a dx^\b\;,
\ee
where we neglected all tidal terms. In the linearised Hubble approximation the metric (\ref{act6}) can be obtained from (\ref{q1}) with the help of special conformal transformation 
\be\la{c4}
y^\a=\O^{-1}(x)\le(x^\a-b^\a x^2\ri)\;,
\ee
where $b^\a=H/2u^\a$, $u^\a=(1,0,0,0)$ is four-velocity of observer in local frame, $\O(x)=1-b_\a x^\a+b^2 x^2$, $b^2=\mm_{\a\b}b^\a b^\b$, and $H=\dot a/a$ is the Hubble constant. All operations of rising and lowering indices are completed with the Minkowski metric. It can be checked that with sufficient accuracy $\O=a(\t)+O(H^2)$.
The reader should notice that transformation (\ref{c4}) yields $a(\t)\eta=x^0+H(x^0)^2/2$ on the worldline of the observer, defined by $x^i=0$. However, $a(\t)\eta=x^0$ for light geodesics, because $x^2=0$ on the light cone. Comparison of these two equations reveals that when $x^0=\t$ on time-like geodesic, the parameter $x^0=\t+H\t^2/2$ on light geodesic, and it does not coincide with the proper time $\t$ of the observer.  

\section{Optical Metric}

The above consideration suggests that equation of light geodesic parametrized with the proper time $\t$ of the observer reads
\be\la{mdu5}
\frac{d^2x^\a}{d\t^2}=H\left(\frac{dx^\a}{d\t}-u^\a\ri)\;.
\ee
The Christoffel symbols entering this equation are not all nil. They can be deduced from the optical metric
\be\la{fe3}
d s^2=-a^2(\t)d\t^2+\d_{ij}dx^i dx^j\;.
\ee
This metric is degenerated in the sense that it is applied only to the events located on the light cone hypersurface connected by the condition $ds=0$. Solution of this equation for a radial propagation of light from/to the origin of the coordinates is
\be\la{tt7}
r=r_0\pm\t+\frac{H}2\t^2\;,
\ee
where $\pm$ sign corresponds to outgoing and incoming light ray respectively, $\t\ge 0$, $r_0$ corresponds to the position of the light-ray particle at the instant of time, $\t=0$. The coordinate speed of light for outgoing ray, $v=dr/d\t=1+H\t$, exceeds the fundamental value $c=1$ for $\t>0$. There is no violation of special relativity here because this effect is non-local. The local value of the speed of light measured at time of emission, $\t=0$, is equal to $c=1$. 

\section{Doppler effect}\la{glb}

A monochromatic electromagnetic wave propagates on a light cone hypersurface of a constant phase $\varphi=\varphi(x^\a)$. The wave vector is $k_\a=\pd_\a\varphi$, and frequency of the wave measured by an observer moving with 4-velocity, $u^\a$, is\cite{2011rcms.book.....K}
$\omega=-k_\a u^\a$.
We denote the point of emission $P_1$, the point of observation $P_2$, and the emitted and observed frequencies as $\omega_1$ and $\omega_2$, respectively. Their ratio $\omega_2/\omega_1=\le(k_\a u^\a\ri)_{P_2}/\le(k_\a u^\a\ri)_{P_1}$
quantifies the Doppler effect. Locally, four-velocity of static observers, $u^\a=(1,0,0,0)$. Hence, 
\be\la{dd2q}
\frac{\omega_2}{\omega_1}=\frac{k_0(\t_2)}{k_0(\t_1)}\;,
\ee
where $\t_1$ and $\t_2$ are the instants of time of emission and observation of light respectively.
The parallel transport of $k_0$ along the light geodesic is computed with the optical metric (\ref{fe3}). Equation of the parallel transport of $k_\a$ in the local coordinates is a light-like geodesic that reads \cite{Synge_GRbook}
\be\la{po1}
\frac{dk_\a}{d\l}=-\frac12 \frac{\pd{g}^{\m\n}}{\pd x^\a}k_\m k_\n\;,
\ee
where $\l$ is the affine parameter along the light ray, the wave vector of light $k^\a=dx^\a/d\l$
Solution of this equation is $k_0(\t)/a(\t)={\rm const.}$. It yields the Doppler shift of frequency of electromagnetic wave in expanding universe as measured in the local inertial frame 
\be\la{d5a}
\frac{\omega_2}{\omega_1}=\frac{a(\t_2)}{a(\t_1)}=1+H(\t_2-\t_1)\;.
\ee
This equation tells us that the cosmological Doppler shift in the local coordinates is {\it blue} ($\o_2>\o_1$) because $\t_2>\t_1$ and $a(\t_2)>a(\t_1)$ due to the Hubble expansion. This may be the reason for the anomalous Doppler shift measured in Pioneer spacecraft mission\cite{kopeikin_2012}.

It is tempting to apply the optical metric formalism to microwave cavity resonator. One might expect the same Doppler drift of the resonator's frequency as given in (\ref{d5a}). At the same time, frequency of atomic clock is not affected by the Hubble expansion\cite{kopeikin_2012}. This type of experiment has been conducted by Storz \etal  \cite{schiller_1998}. They did not find any relative drift between the two frequencies. The reason is that microwave in the cavity is a standing wave that is locked to the boundary of the cavity. As the size of the cavity is not affected by the Hubble expansion locally\cite{kopeikin_2012}, the frequency of the standing wave remains constant. 

\section*{Acknowledgments}
I thank A. Kostelecky for invitation to give a talk at the 6-th Meeting on CPT and Lorentz symmetry, and Department of Physics of the Indiana University (Bloomington) for hospitality. I am grateful to S. Schiller for useful discussions and references.

\end{document}